# A brief introduction of quantum cryptography for engineers

Bing Qi [1,2,*], Li Qian [1,2], Hoi-Kwong Lo [1,2,3,4]

[1]Center for Quantum Information and Quantum Control, University of Toronto,

Toronto, Ontario, Canada

[2]Dept. of Electrical and Computer Engineering, University of Toronto,

Toronto, Ontario, M5S 3G4, Canada

[3]Dept. of Physics, University of Toronto,

Toronto, Ontario, M5S 1A7, Canada

[4]Kavli Institute for Theoretical Physics

Kohn Hall, University of California, Santa Barbara, CA 93106, USA

* bqi@physics.utoronto.ca

# 1 Introduction

On the morning April 18, 1943, Admiral Isoroku Yamamoto, the greatest military commander of the Japanese Navy during World War II, was on his way to a front line base on the island of Bougainville to boost morale. The detailed plan of this visit had been radioed to corresponding units a few days early. All the communications were encoded with the Japanese Naval Cipher JN-25D, an unbreakable code as the Japanese navy believed. But they were wrong. The US naval intelligence unit intercepted and successfully decrypted the message. By the time Yamamoto's plane approached his base, a squadron of P-38 Lightning aircraft from Allies appeared. The rest, as they say, is history. Admiral Yamamoto, the architect for the attack on Pearl Harbor, was defeated by codebreakers.

Cryptography, the art of code-making and code-breaking, plays an important role in human history. It is widely believed that the great feats of Allies' intelligence on routinely breaking enemy's codes, including the famous German Enigma machine, shortened World War II by two years [1]. Nowadays, as Internet and electronic business become more popular, cryptography has also becoming an essential part of our everyday life. With a well developed cryptographic protocol, you can reasonably be sure that all your personal information is well protected whenever you make an online transaction.

The evolution of cryptography has been propelled by the endless war between code-makers and code-breakers, among whom are some of the brightest minds in human history. As soon as an existing code is broken, code-makers need to develop a stronger one to resume secure communication, which in turn stimulates code-breakers to attempt a new attack. The holy grail of cryptography is to develop an absolutely secure coding scheme which is secure against eavesdroppers with unlimited computational power. Surprisingly, this goal was achieved, at least in principle, when Gilbert Vernam invented the one-time pad (OTP) encryption in 1917 [2].

Like in many other modern cryptographic systems, a secure key is employed in the OTP during the encryption and decryption processes. While the encryption algorithm itself is publicly known, the security of the cryptographic system is guaranteed by the security of the key.

As illustrated in Figure 1, OTP is an encryption algorithm where the plaintext (a message understandable to anybody) is encoded with a secret random key (a pad) which has the same length as the plaintext itself. The same key is also used by the legitimate receiver to decode the original message. Given that the random key is only used once, the absolute security of the OTP has been proved by Claude Shannon [3].

Although OTP is unbreakable in principle, there is a problem on applying this scheme in practice: once Alice and Bob have used up their pre-established secure key, the secure communication will be interrupted until they can acquire new key. This is the well known key distribution problem which typically involves two unachievable tasks in classical physics: truly random number generation and unconditionally secure key distribution through an insecure channel. First of all, the deterministic nature of classical physics, which is implied by Albert Einstein's famous quotation "God doesn't play dice", rules out the existence of truly random numbers in chaotic, but classical, processes. In contrast, as we will show in Section 3, truly random numbers can be generated from elementary quantum processes [4]. Secondly, in a world where information is encoded classically, there is no secure scheme to distribute a key through an insecure channel (otherwise, Alice and Bob can employ the same scheme to send secure messages directly). The fundamental reason is that in classical physics, information can be duplicated. Alice and Bob cannot prove that a key established through an insecure channel has not been copied by Eve. The only conceivable but cumbersome way to perform key distribution is by sending trusted couriers. Due to this key distribution problem, OTP has been adopted only when extremely high security is required. Remarkably, even the OTP could be compromised in practice due to inappropriate implementations. For example, during the cold war, KGB spies routinely used the OTP to encode top secret messages. However, some of the random keys had incorrectly been reused by the Soviets. This allowed American and British code-breakers to decrypt some critical messages, including the existence of Soviet espionage at Los Alamos National Laboratories [5]. We will discuss more about the issue of practical security in Section 4.

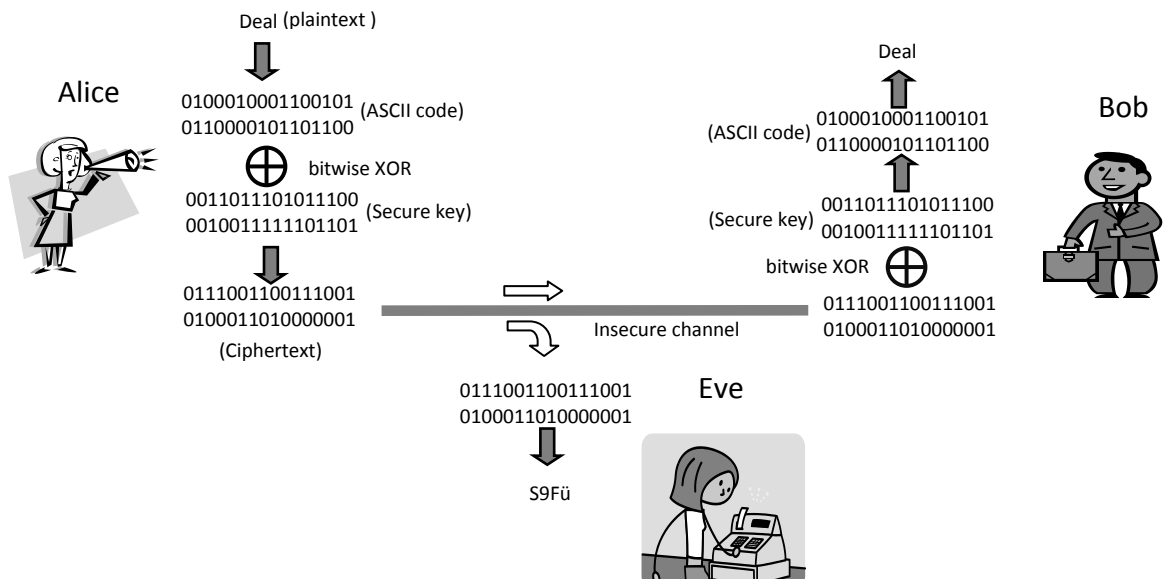

Figure 1 A cartoon illustrates one-time-pad. Alice encodes her message, the ASCII code of "deal", with a secure key by performing bitwise XOR operation. The ciphertext is sent to Bob through an insecure communication channel. Bob can decipher the message by using the same secure key. Although Eve can also acquire a copy of the ciphertext by wiretapping the communication channel, without the secure key she cannot decode the original message.

In most of modern cryptographic systems such as Data Encryption Standard (DES) and Advanced Encryption Standard (AES), much shorter keys are used to encrypt long messages. This reduces the consumption of random keys but does not fully solve the key distribution problem. Furthermore, these protocols are not as secure as the OTP.

To fully solve the key distribution problem, public key cryptographic protocols, including the famous RSA scheme (named after its inventors, Ron Rivest, Adi Shamir, and Leonard Adleman), have been invented [6]. RSA is an asymmetric key algorithm where the message receiver, Bob, prepares two different cryptographic keys—a public key and a private key. Bob broadcasts the public key through an authenticated channel so that everyone who listens to this channel can acquire a copy of the public key. The message sender, Alice, encodes her message with the public key from Bob and sends out the encrypted message through a public insecure channel. This algorithm has been designed in such a way that a message encrypted with a public key can only be decrypted with the corresponding private key.

Public key cryptographic algorithms overcome the key distribution problem and have been widely adopted in today's cryptographic systems. Unfortunately, their security rests upon unproven mathematical assumptions. For example, the security of RSA is based on the assumption that there is no efficient way to find the primes factors of a large integer. However, this assumption has not been proved despite the tremendous efforts from mathematicians. Given the fact that RSA itself was an unexpected discovery, we cannot rule out the possibility that someone could find an efficient factoring algorithm and thus compromise most public cryptographic systems. Moreover, an efficient factoring algorithm running on a quantum computer exists [7]. This suggests that as soon as the first large-scale quantum computer switches on, most of today's cryptographic systems could collapse overnight.

It is true that the realization of a large-scale quantum computer could be still decades away. However, its potential threat to today's information security cannot be neglected. We can imagine that a powerful Eve could record today's communications and decode the information when a quantum computer is available. This is a realistic problem since some information, such as military communications and health records, need to be secure for a long-term.

It is interesting to learn that one decade before people realized that a quantum computer could be used to break public key cryptography, they had already found a solution against this quantum attack—quantum key distribution (QKD) [8]. Based on the fundamental principles in quantum physics, QKD provides an unconditionally secure way to distribute random keys through insecure channels. The secure key generated by QKD could be further applied in the OTP scheme or other encryption algorithms to enhance information security. In this chapter, we will introduce the fundamental principles behind various QKD protocols and present the state-of-the-art QKD technologies. Bearing in mind that the majority of our readers are from engineering and experimental optics, we will focus more on experimental implementations of various QKD protocols rather than on security analysis. Special attention will be given to security studies of real life QKD systems, where unconditional security proofs of theoretical protocols cannot be applied directly due to imperfections. For comprehensive reviews on QKD, see [9-11].

## 2. The principle of quantum cryptography

2.1 Quantum no-cloning theorem

In 1982, N. Herbert proposed a superluminal communication scheme by employing Einstein–Podolsky–Rosen (EPR) pair and by allowing perfectly cloning of an unknown quantum state [12]. This proposal directly conflicts with special relativity and aroused active discussions in the scientific community. Shortly afterwards, W.K.Wootters and W.H. Zurek [13] and D. Dieks [14] independently discovered quantum no-cloning theorem, thus disproving Herbert's superluminal communication scheme.

Quantum no-cloning theorem states that an arbitrary quantum state cannot be duplicated perfectly. This theorem is a direct result of the linearity of quantum physics. Quantum no-cloning theorem is closely related to another important theorem in quantum mechanics, which states: if a measurement allows one to gain information about the state of a quantum system, then in general the state of this quantum system will be disturbed, unless we know in advance that the possible states of the original quantum system are orthogonal to each other.

Instead of providing a mathematical proof of quantum no-cloning theorem, we simply discuss two examples to show how it works. In the first case, we are given a photon whose polarization is either vertical or horizontal. To determine its polarization state, we can send it through a polarization beam splitter followed by two single photon detectors. If the detector at the reflection path clicks, we know the input photon is vertically polarized, otherwise it is horizontally polarized. Once we know the polarization state of the input photon, we can prepare arbitrary number of photons in the same polarization state. Equivalently, we have achieved perfect cloning of the polarization state of the photon. This is because the two possible polarization states of the input photon are orthogonal to each other. In the second case, we are given a photon whose polarization is randomly chosen from a set of {horizontal, vertical, 45°, 135°}. Since the four polarization states given above are linearly dependent, it is impossible to determine its polarization state from any experiment. For example, if we use the same polarization beam splitter mentioned above, a 45° polarized photon will have a 50/50 chance to be either reflected or transmitted, therefore it cannot be determined with certainty. Note in this chapter, we simply treat a photon as an electromagnetic wave package with an undividable energy of hν, where h is the Planck constant and ν is the frequency of light.

One common question is why an optical amplifier, which has been widely used in optical communication to boost optical power, cannot be used to copy photons. Actually, in the original superluminal communication scheme, Herbert did mistakenly assume that the receiver could make perfect copies of the input photon with an optical amplifier [12]. As pointed out by Wootters, the impossibility of making perfect copies of photon through stimulated emission process originates from the unavoidable spontaneous emission: while the stimulated photon is a perfect copy of the incoming one, the spontaneous emitted photon has a random polarization state [13].

At first sight, the impossibility of making perfect copies of unknown quantum states seems to be a shortcoming. Surprisingly, it can also be an advantage. It turned out that by using this impossibility smartly, unconditionally secure key distribution could be achieved: any attempts by the eavesdropper to learn the information encoded quantum mechanically will disturb the quantum state and expose her existence. In next section, we will present the well known Bennett-Brassard-1984 (BB84) QKD protocol.

2.2 The BB84 quantum key distribution protocol

The first attempt of using quantum mechanics to achieve missions impossible in classical information started in the early 70's. Stephen Wiesner, a graduate student at Columbia University, proposed two communication modalities not allowed by classical physics: "quantum multiplexing" channel (this term followed from [15]) and counterfeit-free bank-note. Unfortunately, his paper was rejected and couldn't be published until a decade later [16]. In 1980's, Charles H. Bennett and Gilles Brassard extended Wiesner's idea and applied it to solve the key distribution problem in classical cryptography. In 1984, the well known BB84 QKD protocol was published [8].

The basic idea of the BB84 protocol is surprisingly simple. Let us assume that both Alice and Bob stay in highly secured laboratories which are connected by an insecure communication channel, for example, an optical fiber. Eve is allowed to fully control the communication channel but she is not allowed to sneak into Alice or Bob's laboratory to steal information directly. Furthermore, Alice and Bob share an authenticated public channel (a channel that Eve can listen but she cannot interfere) and they can generate random numbers locally. The question is how they can establish a long string of shared random numbers as secure key.

Being aware of quantum no-cloning theorem, Alice encodes her random bits on the polarization state of single photons and sends them to Bob through an insecure quantum channel. Here, we use the term "quantum channel" to emphasis the fact that information through this channel is encoded on the quantum state of photons.

As the first try, she uses a horizontally polarized photon to encode bit "0" and a vertically polarized photon to encode bit "1". Bob can decode the random bit by performing polarization measurement. However, this scheme turns out to be insecure: when Alice's photon travels through the channel, Eve can intercept it and measure its polarization state. After that Eve can prepare a new photon according to her measurement result and send it to Bob. By the end of the day, Eve has a perfect copy of whatever Bob has. No secure key can be generated. Then Alice realizes what the problem is: the quantum no-cloning theorem cannot be applied to a set of orthogonal states. After some thinking, she introduces a new concept—"basis", which represents how the random bits are encoded. In basis one (rectilinear basis), she uses horizontal polarization to represent bit "0" and vertical polarization to represent bit "1". In basis two (diagonal basis), she uses 45° polarization to represent bit "0" and 135° polarization to represent bit "1". For each transmission, Alice randomly chooses to use either rectilinear or diagonal basis to encode her random number. Now, the polarization of each photon is randomly chosen from a set of {horizontal, vertical, 45°, 135°} and it is impossible for Eve to determine its polarization state. If Eve uses a polarization beam splitter to project the input photon into either horizontal or vertical polarization state (we call it a measurement in rectilinear basis), then she will destroy information encoded in diagonal basis, since a 45° or 135° polarized photon has the same chance to be projected into either horizontal or vertical polarization state. Eve is puzzled, so does Bob, because neither of them knows which basis Alice will choose in advance.

Without the knowledge of Alice's basis selection, Bob randomly chooses either rectilinear or diagonal basis to measure each incoming photon. If Alice and Bob happen to use the same basis, they can generate correlated random bits. On the other hand, if they use different basis, their bit values are uncorrelated. After Bob has measured all the photons, he compares his measurement bases with Alice through an authenticated public channel. They only keep random bits generated with matched bases, which are named as sifted keys. Without environmental noises, system imperfections and Eve's disturbance, their sifted keys are identical and can be used as a secure key.

The existence of an authenticated channel between Alice and Bob, which gives Bob an advantage over Eve, is essential to the security of QKD. Otherwise, Eve can always break the security by performing man-in-the-middle attack: she can pretend to be Bob and communicates with Alice; in the mean time, she can pretend to be Alice and communicates with Bob. Under this condition, no secure key can be generated. The authentication can be assured if Alice and Bob share a short secure key in advance. In this sense, a QKD protocol is actually a quantum key expansion protocol: it takes a short shared key and expands it into an information-theoretically secure long shared key [17]. The universal composability of QKD implies that we can use part of the secure key generated in one round of QKD for the authentication purpose of the next round of QKD.

Let us see what will happen if Eve launches a simple "intercept and resend" attack: for each photon from Alice, Eve performs a measurement in a randomly chosen basis and resends a new photon to Bob according her measurement result. Let us focus on those cases when Alice and Bob happen to use the same bases since they will throw away other cases anyway. If Eve happens to use the right basis, then both she and Bob will decode Alice's bit value correctly. On the other hand, if Eve uses the wrong basis, then both she and Bob will have random measurement results. This suggests that if Alice and Bob compare a subset of the sifted key, they will see a significant amount of errors. It can be easily shown that the intercept and resend attack will introduce a 25% quantum bit error rate (QBER). This example illustrates the basic principle behind QKD: *Eve can only gain information at the cost of introducing errors which will expose her existence*.

In practice, quantum bit errors can be originated from either the intrinsic noises of the QKD system or Eve's attack, so Alice's sifted key is typically different from Bob's, and Eve may have partial information about it. To establish perfectly correlated key between Alice and Bob, and remove Eve's knowledge about the final key, error correction and privacy amplification algorithms can be applied: Alice and Bob estimate the information gained by Eve during the quantum transmission stage from the observed QBER and some parameters of the QKD system. They also estimate the information leakage during the error correction process. If Eve's information is too much, no secure key can be generated and they have to restart the whole QKD protocol. On the other hand, if Eve's information is below certain threshold, Alice and Bob can perform privacy amplification algorithm to generate a shortened final key on which Eve's information is exponentially small.

Typically, a QKD protocol can be separated into a quantum transmission stage and a classical post-processing stage. The former includes quantum state preparation, transmission and detection, while the latter includes bases comparison, error correction and privacy amplification. The procedures of the BB84 protocol are summarized in Table 1.

A few remarks about the implementation of the BB84 QKD protocol:

First of all, any communication channel (such as telecom fiber) has loss and the efficiencies of practical single photon detectors are typically low. This suggests that a significant portion of photons sent by Alice will not be registered by Bob. Since Alice's photons only carry random bits but not useful message, this loss-problem can be easily resolved by employing a post-selection process: Bob publicly announces which photons he registers and Alice only keeps the corresponding data.

Secondly, any practical QKD system has intrinsic QBER, which could originate from noises of detectors, imperfections of polarization control system, background noises, etc. The total observed QBER contains both the intrinsic QBER and the QBER due to Eve's attack. The question is: can Alice and Bob distinguish these two types of errors? Most QKD security proofs adopt a conservative assumption that all the observed errors are due to Eve. Under this assumption, Eve's information gain could be overestimated which results in a lower secure key rate.

Table 1 Summary of the BB84 QKD protocol

| Alice's random bits | 1 | 0 | 0 | 1 | 0 | 1 | 1 | 1 | 0 | 1 |
|---|---|---|---|---|---|---|---|---|---|---|
| Alice's encoding bases | × | × | + | × | + | + | + | × | + | + |
| Alice's photon polarization | 135° | 45° | H | 135° | H | V | V | 135° | H | V |
| Bob's measurement bases | + | × | + | × | × | + | × | + | + | + |
| Bob's measurement result | H | 45° | H | * | 135° | V | 45° | * | H | V |
| Bob's raw data | 0 | 0 | 0 | | 1 | 1 | 0 | | 0 | 1 |
| Sifted key from matched bases | | 0 | 0 | | | 1 | | | 0 | 1 |
| +: rectilinear basis; ×: diagonal basis; H: horizontal polarization; V: vertical polarization; *: no detection ||||||||||| |

The basic idea of the BB84 QKD protocol is beautiful and its security can be intuitively understood from the quantum no-cloning theorem. On the other hand, to apply QKD in practice, Alice and Bob need to find the upper bound of Eve's information quantitatively, given the observed QBER and other system parameters. This is the primary goal of various QKD security proofs and it had turned out to be extremely difficult. One major challenge comes from the fact that Eve could launch attacks way beyond today's technologies and our imaginations. Nevertheless, more than a decade after its invention, QKD was proved to be unconditionally secure [18-20]. This is one of the most significant achievements in quantum information.

All QKD security proofs are based on certain assumptions of the QKD system. Some security proofs have been built on "realistic QKD models" by taking into account of practical imperfections, such as channel loss, detector noise, etc [21]. Nevertheless, it is impossible to incorporate all the imperfections of a real-life QKD system into a theoretical model, even in principle. It is natural to ask the following question: is it possible that a small neglected imperfection spoils the security of the whole system? We will come back to this question in Section 4.

The BB84 protocol had been ignored by almost the whole scientific community until early 90's when its founders decided to build a prototype. In October 1989, the first photon carried a random bit travelled over a distance of 32.5cm in air [22]. As Brassard recalled in a lively essay [15], "It remains a mystery to me that this successful prototype made a world of difference to physicists, who suddenly paid attention". Since then, significant progresses have been achieved and even commercial QKD systems are available. We will present more details about the recent progresses in Section 3.

2.3 Entanglement based QKD

Since Einstein, Podolsky and Rosen published the well known "EPR" paper in 1935 [23], entanglement has been one of most puzzling yet attractive feature in quantum mechanics. To illustrate how entanglement can be used to achieve secure key distribution, let us first discuss some properties of a polarization-entangled photon pair.

An arbitrary polarization state of a single photon can be described by a superposition of two basis states

$$|\psi\rangle_s = \alpha|\updownarrow\rangle + \beta|\leftrightarrow\rangle \qquad (1)$$

where $|\updownarrow\rangle$ and $|\leftrightarrow\rangle$ represent vertical and horizontal polarization states, which constitute a set of orthogonal bases. $\alpha$ and $\beta$ are complex numbers satisfying the normalization condition $\alpha\alpha^* + \beta\beta^* = 1$.

In writing down equation (1), we have assumed that the polarization of photon is described by a pure state—there is a well defined phase relation between the two basis components. In contrast, a non-polarized photon is in a mixed state which can only be interpreted as a statistic mixture of basis states.

Similarly, the most general polarization state (pure state) of a photon pair can be described by a superposition of four basis states

$$|\psi\rangle_{pair} = \alpha_1|\updownarrow\rangle_1|\updownarrow\rangle_2 + \alpha_2|\updownarrow\rangle_1|\leftrightarrow\rangle_2 + \alpha_3|\leftrightarrow\rangle_1|\updownarrow\rangle_2 + \alpha_4|\leftrightarrow\rangle_1|\leftrightarrow\rangle_2 \qquad (2)$$

Here $|\updownarrow\rangle_1|\updownarrow\rangle_2$ represents a basis state that both photons are in vertical polarization state. The other three terms in equation (2) can be understood in a similar way. In the special case when $\alpha_1 = \alpha_4 = 1/\sqrt{2}$ and $\alpha_2 = \alpha_3 = 0$, we have one type of polarization entangled EPR photon pair

$$|\Phi\rangle_{pair} = 1/\sqrt{2}\left(|\updownarrow\rangle_1|\updownarrow\rangle_2 + |\leftrightarrow\rangle_1|\leftrightarrow\rangle_2\right) \qquad (3)$$

One special feature of the above state is that it cannot be described by a tensor product $|\Phi\rangle_{pair} \neq |\psi\rangle_1 \otimes |\psi\rangle_2$, where $|\psi\rangle_1$ and $|\psi\rangle_2$ are arbitrary single photon polarization states. In other words, the two photons are "entangled" with each other. Entangled photons can present non-local correlation which doesn't exist in classical physics.

Suppose we send one photon of an EPR pair to Alice and the other one to Bob. If Alice measures her photon in the rectilinear basis, she will detect either a vertical or a horizontal polarized photon with the same probability. Depending on Alice's measurement result, Bob's photon will be projected to the corresponding polarization state. If Bob subsequently measures his photon in the same basis, his measurement result will be perfectly correlated to Alice's result. On the other hand, if Bob measures in diagonal basis, no correlation exists. The above arguments are also applicable if Bob performs his measurement first.

What happens if both Alice and Bob measure in the diagonal basis? Surprisingly, they still get perfect correlation. This can be shown by rewriting equation (3) into the following form

$$|\Phi\rangle_{pair} = \frac{1}{\sqrt{2}}\left(|+\rangle_1|+\rangle_2 + |-\rangle_1|-\rangle_2\right) \quad (4)$$

where $|+\rangle = \frac{1}{\sqrt{2}}(|\updownarrow\rangle + |\leftrightarrow\rangle)$ represents 45° polarization state and $|-\rangle = \frac{1}{\sqrt{2}}(|\updownarrow\rangle - |\leftrightarrow\rangle)$ represents 135° polarization state.

The discussion above suggests that Alice and Bob can implement BB84 type QKD based on an EPR source, as shown in Figure 2. The EPR source can be placed between Alice and Bob. One photon of each EPR pair is sent to Alice and the other one to Bob. For each incoming photon, Alice and Bob randomly and independently choose their measurement bases to be either rectilinear or diagonal. After they have measured all the photon pairs, Alice and Bob will compare their measurement bases and only keep random bits generated with matched bases. Similar to the BB84 QKD based on a single photon source, they can further perform error correction and privacy amplification to generate the final secure key. Note that, before Alice and Bob perform the measurement, the polarization of each photon is undetermined. More precisely, each photon of the EPR pair is in a maximum mixed state (fully non-polarized). The eavesdropper cannot gain any information from the photon when it transmits from the EPR source to the user, because there in no information encoded. The random bits are generated during the measurement processes.

The above QKD protocol based on EPR pairs is quite similar to the original BB84 QKD. However, there is much more physical insight in the first entanglement based QKD protocol proposed by Ekert in 1991[24], especially the deep connection between entanglement and the security of QKD. In his original proposal, Ekert suggested that Alice and Bob can verify entanglement by testing a certain type of Bell's inequalities [25]. As long as they can verify the existence of entanglement, it is possible to generate secure key.

Without discussing more details about Bell's inequality, we simply remark that Alice and Bob can perform Bell's inequalities test without knowing how the measurement results are acquired. This has inspired the so called "device independent" security proof, which will be discussed in Section 4.

Entanglement based QKD could yield better security under certain circumstances. Furthermore, the long distance QKD scheme based on quantum repeaters, which will be discussed in Section 3, can only be implemented with entanglement based QKD protocols. However, the technical challenges on implementing entangled QKD are much bigger than that of the BB84 QKD. Readers interested in this topic can find more details in [9].

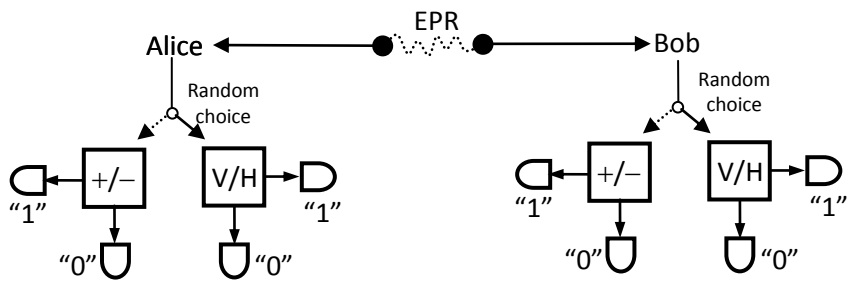

Figure 2  Entanglement based QKD. Each of Alice and Bob measures half of an EPR pair in a randomly chosen basis. V/H: rectilinear basis; +/−: diagonal basis.

## 2.4 Continuous variable QKD

In the BB84 QKD protocol, Alice's random bits are encoded in a two dimensional space like the polarization state of a single photon. More recently, QKD protocols working with continuous variables have been proposed. Among them, the Gaussian modulated coherent state (GMCS) QKD protocol has drawn special attention [26].

The basic scheme of the GMCS QKD protocol is shown in Figure 3: Alice modulates both the amplitude quadrature and phase quadrature of a coherent state with Gaussian distributed random numbers. In classical electromagnetism, these two quadratures correspond to the in-phase and out-of-phase components of electric field, which can be conveniently modulated with optical phase and amplitude modulators. Alice sends the modulated coherent state together with a strong local oscillator (a strong laser pulse which serves as a phase reference) to Bob. Bob randomly measures one of the two quadratures with a phase modulator and a homodyne detector. After performing his measurements, Bob informs Alice which quadrature he actually measures for each pulse and Alice drops the irrelevant data. At this stage, they share a set of correlated Gaussian variables which are called the "raw key". Given the variances of the measurement results below certain thresholds, they can further work out perfectly correlated secure key by performing reconciliation and privacy amplification.

The security of the GMCS QKD can be comprehended from the uncertainty principle. In quantum optics, the amplitude quadrature and phase quadrature of a coherent state form a pair of conjugate variables, which cannot be simultaneously determined with arbitrarily high accuracies due to Heisenberg uncertainty principle. From the observed variance in one quadrature, Alice and Bob can upper bound Eve's information about the other quadrature. This provides a way to verify the security of the generated key. Recently, an unconditional security proof of the GMCS QKD appeared [27].

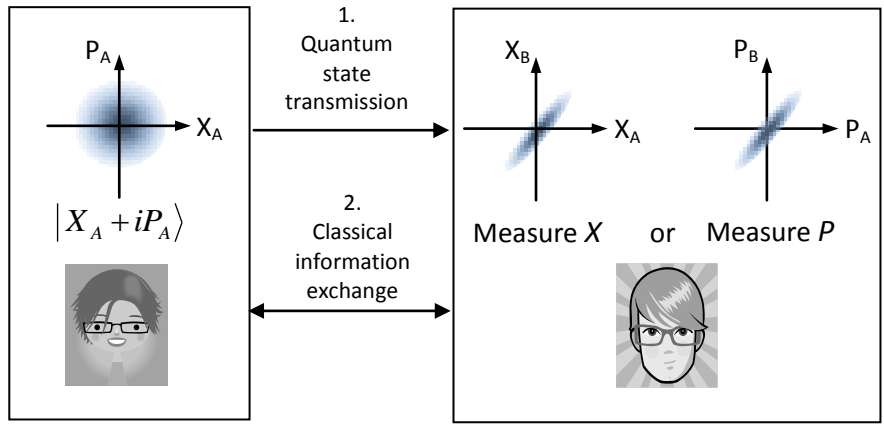

Figure 3 The Gaussian modulated coherent state (GMCS) QKD. *X*: amplitude quadrature; *P*: phase quadrature.

Different from the BB84 QKD, in GMCS QKD, homodyne detectors are employed to measure electric fields rather than photon energy. By using a strong local oscillator, high efficiency and fast photo diodes can be used to construct the homodyne detector which could result in a high secure key generation rate. However, the performance of the GMCS QKD is strongly dependent on the channel loss. Recall that in the BB84 QKD system, the channel loss plays a simple role: it reduces the communication efficiency but it will not introduce QBER. A photon is either lost in the channel, in which case Bob will not register anything, or it will reach Bob's detector intact. On the other hand, in the GMCS QKD, the channel loss will introduce vacuum noise and reduce the correlation between Alice and Bob's data. As the channel loss increases, the vacuum noise will become so high that it is impossible for Alice and Bob to resolve a small excess noise (which is used to upper bound Eve's information) on the top of a huge vacuum noise.

Comparing with the BB84 QKD, the GMCS QKD could yield a high secure key rate over short distances [28, 29].

## 3. State of the art QKD technologies

A QKD system is built upon cutting edge technologies, such as single photon detector, truly random number generator, EPR photon source and phase stabilized interferometer, etc. In pursuit of higher secure key rate and longer key distribution distance, many conventional techniques have been significantly improved and new tools have been developed. In this section, we will present state of the art QKD technologies.

3.1 Sources for QKD

(1) Single photon source

Although the term "single photon" has been used extensively in quantum optics and quantum communication, devices generating individual photons on demand are still not ready for practical applications despite the significant progresses achieved in the last decade.

A perfect single photon source is sometimes called a "photon gun": each time Alice fires, a light pulse containing exactly one photon will be emitted. The single-photon state is a special "photon-number" state—a state contains a precise number of photons. The conventional laser source, regardless how stable it is, cannot produce photon number states. This can be intuitively understood as follows: the gain medium of a conventional laser contains a large number of individual atoms or molecules and each of them has a certain probability to emit a photon during a fixed time period. Thus a statistical fluctuation of the total photon number is expected. To generate a single photon state, it is more likely that a microscopic source containing a single emitting object should be employed: the single emitter can be pumped from its ground state to an excited state during an excitation cycle; after that, a single photon could be generated through spontaneous emission. So far, various single emitters have been investigated for single photon generation, including atoms or ions in gas phase, organic molecules, colour centres in crystals, semiconductor nanocrystals, quantum dots, etc. For a comprehensive review, see [30].

(2) EPR photon pair

Entangled photon pairs can be generated through nonlinear optical processes, such as spontaneous parametric down-conversion (SPDC) [31]. In this process, a pump photon spontaneously decays into a pair of daughter photons in a nonlinear crystal. The conservations of energy and momentum imply that the generated daughter photons are entangled in spectral and spatial domains. Like other phase sensitive nonlinear processes, SPDC requires phase matching to be efficient. The phase matching means that a proper phase relationship between pump light and down-converted light should be maintained throughout the nonlinear crystal, so the probability amplitudes of the SPDC process at different locations will add up coherently. The phase matching requirement and the birefringence of nonlinear crystal suggest that it is possible to selectively generate photon pairs with certain polarization state. This is because down-converted photons in other polarization states cannot satisfy phase matching condition and thus cannot be produced efficiently.

One way to generate polarization entangled photon pairs is as follows [31]. We can place two different nonlinear crystals together: the first crystal can be selected to phase match vertical pump photon with horizontal down-converted photon pair; while the second crystal can be chosen to phase match horizontal pump photon with vertical down-converted photon pair. With carefully designed geometry, a 45° polarized pump photon will be equally likely to down convert in either crystal and these two possible down-conversion processes add coherently. Thus high quality polarization entangled photon can be generated.

Entangled photon pairs can also be generated in other nonlinear processes, such as four-wave mixing in optical fiber [32]. We will not discuss more details but simply remark that nonlinear optics has been playing an important role in both classical optical communication and quantum communication. The knowledge we have accumulated in classical domain could be the most valuable resource as we venture into the quantum domain.

(3) Attenuated laser source and decoy-state QKD

In the original BB84 QKD protocol, Alice encodes her random bits on single photon states. However, as we mentioned earlier, an efficient single photon source with high quality is not available in practice. Instead, heavily attenuated laser sources have been employed in most practical QKD systems. Obviously, this substitution has to be taken into account in security analysis.

A heavily attenuated laser pulse can be modeled as a coherent state which is a superposition of different photon number states. In practice, we can decrease the average photon number of a laser pulse by sending it through an optical attenuator. However, no matter how small the average photon number is, there is always a non-zero probability that the laser pulse contains more than one photon. This opens a door for the so called photon number splitting (PNS) attack [33]. In PNS attack, Eve performs a special "quantum non-demolition" (QND) measurement to learn the photon number information of Alice's laser pulse without destroying it or disturbing the encoded quantum information. If the laser pulse contains one photon, Eve simply blocks it and Bob will not receive anything. On the other hand, if the laser pulse contains more than one photon, Eve splits out one photon and sends the rest to Bob through a lossless channel. Eve stores the intercepted photons in quantum memories until Bob announces his measurement bases. Then she measures her photons in the same bases as Bob. In the end, Eve has an exact copy of whatever Bob has. No secure key can be generated.

Eventually, a security proof for the BB84 QKD based on "realistic model" appeared [21]. Stimulated by the PNS attack, one crucial argument used in the above proof is: among all of Bob's detection events, only those originated from single photon pulses sent by Alice contribute to the secure key. In other words, all the multi-photon pulses in BB84 are insecure. The multi-photon probability of a coherent state can be effectively suppressed by applying heavy attenuation to lower the average photon number; however, this also significantly increases the proportion of vacuum state, which in turn lowers the efficiency. As a result, the final secure key rate scales quadratically with the transmittance of the quantum channel. This is much worse than a QKD system based on single photon source, where a linear relation between the secure key rate and the transmittance is expected.

One breakthrough came in 2003 when the "decoy" idea was proposed [34]. Recall that in the PNS attack, Eve selectively blocks single photon pulses. The resulting quantum channel has a photon number dependent transmittance, which is much different from a passive channel. The insight of the decoy idea is that the PNS attack can be detected by testing the quantum channel during QKD process.

In classical metrology, one common way to calibrate an unknown device is to test its responses at difference input signals. The same strategy has been adopted in the decoy-state QKD: Alice and Bob conduct QKD with laser pulses having different average photon numbers (which are named as either "signal state" or "decoy states") and evaluate their transmittances and QBERs separately. A PNS attack by Eve will inevitably result in different transmittances for signal state and decoy states and thus can be detected. It has been shown that the secure key rate of the decoy-state BB84 QKD scales linearly with the transmittance of the quantum channel, which is comparable to the secure key rate of the BB84 QKD with a perfect single photon source.

One crucial assumption in the decoy-state QKD is that the signal state and decoy states are identical except for their average photon numbers. This means after Eve's photon-number measurement, she has no way of telling whether the resulted photon number state is originated from the signal state or decoy states. The unconditional security of the decoy-state BB84 QKD was given in 2005 [35]. Practical decoy-state protocols were proposed in [36-38], and the experimental implementation appeared in 2006 [39]. Today, the decoy-state idea has been widely adopted in BB84 QKD systems implemented with weak coherent state sources [40-43].

3.2 Quantum state detection

The intrinsic noises of a QKD system operating through a high loss-channel are mainly determined by the noises of Bob's detectors. Here, we will present two technologies for quantum state detection which have been applied in QKD systems.

(1) Single photon detector

In the BB84 QKD protocol, Bob measures Alice's photons with single photon detectors (SPD). Conventional photon counters, such as photomultiplier tubes (PMT) and avalanche photodiodes (APD) are typically operated in highly nonlinear region: a single incoming photon can generate a macroscopic electrical pulse through multiplication processes. This suggests that these detectors are threshold detectors: they can distinguish the existence of a photon from the vacuum state, but cannot discriminate the actual photon number. More recently, superconductive single photon detectors (SSPD) have been invented and demonstrated superior performances [44, 45].

The most important parameters of a SPD include detection efficiency, dark count probability, dead time and time jitter. (a) Detection efficiency is the probability that a SPD clicks upon receiving an incoming photon. The overall detection efficiency is determined by both intrinsic quantum efficiency and coupling losses of the detector. Typically, the detection efficiency is a function of the wavelength of the incoming photon and it may also be polarization sensitive. (b) Dark counts are detection events registered by a SPD while no actual photon hits it (i.e. "false alarm"). Due to its random nature, the QBER contributed by dark count events is 50%. The maximum key distribution distance of a QKD system is typically determined by the dark count probability of the SPD. (c) Dead time is the recovery time of a SPD after it registers a photon. During this period, the SPD will not respond to input photons. (d) Time jitter is the random fluctuation of a SPD's output electrical pulse in time domain.

The performances of two commonly encountered SPDs are summarized below:

**APD-SPD**

APDs operated in Geiger mode have been commonly used in QKD as single photon detectors [46]. In Geiger mode, the applied bias voltage on an APD exceeds its breakdown voltage and a macroscopic avalanche current can be triggered by a single absorbed photon.

In free space QKD systems, which are typically operated at the wavelength range of 700~800nm, silicon APDs are the natural choice and high performance commercial products are available. The overall efficiency of a silicon APD-SPD can be above 60% with a dark count rate less than 100 counts per second. The time jitter can be as small as 50ps and the typical dead time is in the range of tens of ns [47]. Determined by its dead time, the maximum count rate of a silicon APD-SPD is about a few tens of MHz.

In QKD systems operated at telecom wavelengths, InGaAs/InP APDs have been routinely applied [48]. Comparing with silicon APDs, InGaAs/InP APDs operated in Geiger mode have particularly high afterpulsing probabilities: a small fraction of charge carries generated in previous avalanche process can be trapped in the APD and trigger additional "false" counts later on. These false counts are equivalent to signal-dependent "dark counts". To reduce the afterpulsing probability, the gated Geiger-mode has been introduced: most of the time, the bias voltage on an APD is below its breakdown voltage thus the APD will not response to incoming photons; only when Alice's laser pulses are expected, the bias voltage is raised above the breakdown voltage for a short time window to activate the SPD. After the SPD registers a photon, the bias voltage will be kept below the breakdown voltage for a long time (typically, a few microseconds) to release the trapped carriers completely. This is equivalent to introduce an external "dead time" which significantly limits the operating rate of this type of SPD. Typically, an InGaAs/InP SPD has an efficiency of 10%, a dark count rate of $10^{-6}$ per nanosecond and a maximum count rate less than 1MHz. Recently, the performance of the gated InGaAs/InP APD has been improved significantly by using either self-differential scheme [49] or sinusoidal gating scheme [50]. The idea is to improve the sensitivity of discrimination circuit so that the APD can be operated with a relatively low avalanche gain. This results in a much smaller afterpulsing probability. These SPDs have been successfully applied in QKD systems operated at GHz pulse repetition rates.

Another interesting idea to detect single photons at telecom wavelength is to "up-shift" its frequency by employing parametric frequency up-conversion process and then detects the up-converted photon with a silicon APD [51]. However, this approach suffers from the high noise accompanying with the frequency up-conversion process. Furthermore, these detectors have quite narrow spectral response range and typically are polarization sensitive.

**Superconductive SPD**

Single photon detectors exploiting superconductivities, such as superconducting SPD (SSPD) [44] and transition edge sensors (TES) [45], have been successfully developed. Both have wide spectral response range covering the visible and telecom wavelengths. The TES has both a high intrinsic efficiency and an extremely low dark count rate. However, due to its long relaxation time, it cannot operate above a few MHz. SSPD, on the other hand, has shown a very small relaxation time and has been employed in a QKD system operated at 10GHz repetition rate [52]. Currently, these detectors have to be operated at temperatures around a few Kelvin which may not be convenient for practical applications.

(2) Optical homodyne detector

In principle, the homodyne detectors used in the GMCS QKD have no difference from the one used in classical coherent communication system. In either case, the sender encodes information on either the in-phase or quadrature component of electric field (in terms of quantum optics, the information is encoded on either the amplitude quadrature or phase quadrature of a coherent state), while the receiver decodes the information by interfering the signal with a local oscillator. However, the performance of a GMCS QKD is much more sensitive to electric noises of the homodyne detector. To effectively detect Eve's attack, one basic requirement is that the homodyne detector has to be shot noise limited, which suggests that a strong local oscillator is required.

Comparing with an SPD, a homodyne detector normally has a higher efficiency and a larger bandwidth. This is due to the fact that a strong local oscillator is used to produce interference with the weak quantum signal during homodyne detection. The resulting interference signals are strong enough to be detected with high speed photo diodes. However, there are also great challenges to the implementation of homodyne detection in QKD. First of all, to stabilize the relative phase between the signal and the local oscillator, both of them are normally split from the same laser pulse and propagate from Alice to Bob through the same optical fiber. Since the local oscillator is much stronger than the quantum signal, special attention is required to separate them effectively on the receiver's side. Secondly, in a GMCS QKD, a balanced detection scheme is commonly employed to remove the DC background. In practice, a small asymmetry between the two detection channels could spoil the performance of the whole system.

### 3.3 Quantum random number generator

Quantum random number generator (QRNG) is both a critical device in most QKD systems and a standalone product by itself offered by quantum physics. In a "prepare & measure" QKD system, such as the BB84 QKD, Alice needs truly random numbers for selecting encoding bases and bit values, while Bob may also need random numbers for choosing his measurement bases. The only exception is the entanglement based QKD protocol implemented with passive bases selection scheme. In this special case, the bases selection can be accomplished passively with a beam splitter and the random bits are generated during measurement processes. However, even in this case, truly random numbers may be still required during classical post-processing stage, for example, privacy amplification with random hashing.

Conventional pseudorandom number generators based on algorithms or physical random number generators based on chaotic behaviors of complex systems are not suitable for this application. Given the seeds of a pseudorandom number generator or the initial condition of a classical chaotic system, their future outputs are completely predictable. Actually, the deterministic nature of classical physics rules out the existence of truly random numbers. In contrast, the probabilistic nature of quantum mechanics suggests that truly random numbers could be generated from elementary quantum processes [4].

Most of today's QRNGs are implemented at single photon level. The basic principle can be easily understood from Figure 4: single photon pulses are sent to a symmetric beam splitter and two SPDs are used to detect transmitted and reflected photon, respectively. Depending on which SPD clicks, the registered event is assigned as either bit "1" or bit "0". A QRNG with a random number generation rate of 16 Mbit/sec is commercial available (id Quantique, Switzerland).

Another promising quantum random number generation scheme is based on measuring random field fluctuation of vacuum [53], or literally speaking, to generate random numbers from "nothing". This scheme can be implemented by sending a strong laser pulse through a symmetric beam splitter and detecting the differential signal of the two output beams with a balanced receiver. To acquire high quality random numbers, the electrical noise of the balanced receiver has to be orders of magnitude lower than the shot noise.

Most recently, a high-speed QRNG based on measuring the quantum phase noise of a single-mode semiconductor laser has been proposed and a 500Mbit/sec random number generation rate has been demonstrated [54].

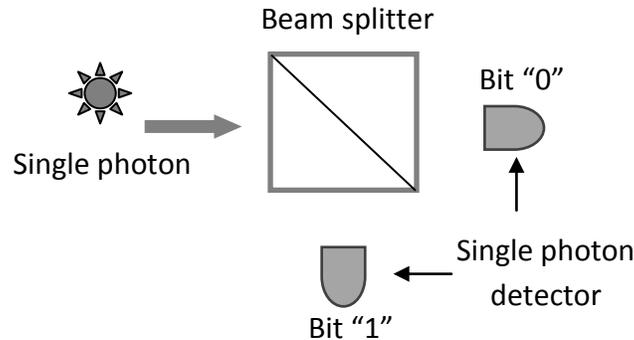

Figure 4 The basic principle of a QRNG based on sending a single photon through a beam splitter and observing which path it emerges from.

3.4 Point-to-point QKD

(1) QKD experiments through telecom fiber

The great achievements of classical fiber optical communication and the availability of a worldwide fiber network suggest that single mode fiber (SMF) could be the best choice as the quantum channel for practical QKD systems.

There are two basic requirements on the quantum channel: low loss and weak decoherence. The first requirement is obvious. The secure key can only be generated from photons detected by Bob. Any channel loss will lower the efficiency of the QKD system. The second requirement, weak decoherence, means that the disturbance of the quantum state by the channel should be as small as possible.

While the loss of standard SMF are relatively stable, the decoherence introduced by a long fiber link is sensitive to both the encoding method and the environment noises. Owing to the residual birefringence in optical fiber, the polarization of a photon will change as it propagates through a long fiber. If the polarization distortion due to the fiber link is time-invariant, Alice and Bob can compensate this effect by calibrating the channel in advance. However, in practice, the birefringence of a SMF is sensitive to the environmental noises and it can be very challenging to compensate a fast polarization fluctuation. This is one of the reasons why most of fiber based QKD systems adopt phase-coding scheme: Alice encodes her random bit on the relative phase between two laser pulses and sends both of them to Bob through a quantum channel while Bob decodes the phase information by performing interference experiment. It can be shown that in principle a phase-coding BB84 QKD is equivalent to a polarization-coding BB84 QKD. Recently, a phase-coding BB84 QKD system incorporating decoy states has been implemented at GHz operating rate [55]. The achieved secure key rate is about 1Mbit/sec over a 20-km fiber link and 10kbit/sec over a 100-km fiber link. Several companies have brought QKD into the commercial market, including id Quantique in Switzerland, MagiQ in USA and SmartQuantum in France. Figure 5 shows a Cerberis QKD system from id Quantique.

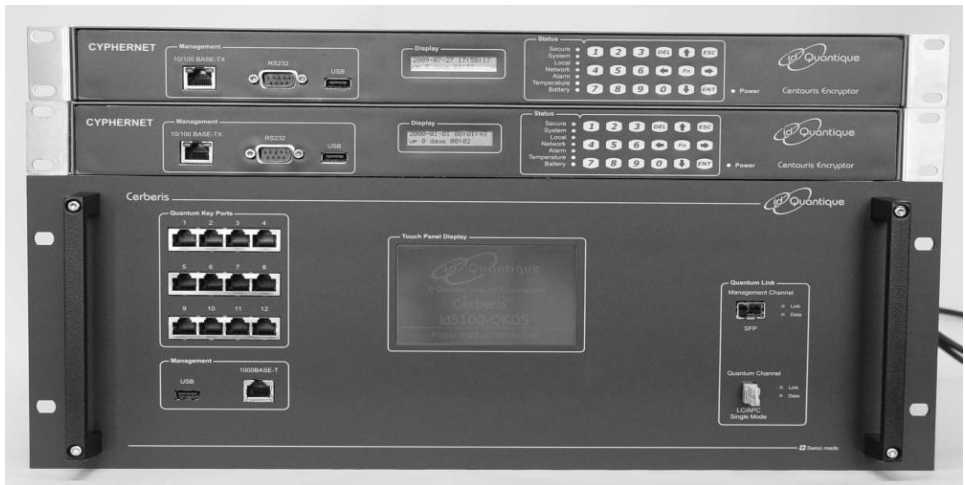

Figure 5 Commercial QKD system fabricated by id Quantique (this picture is used with permission of id Quantique)

It is worthwhile to mention a special "plug & play" QKD configuration which has been adopted in many commercial QKD systems [56]. As suggested by its name, the "plug & play" QKD configuration has been demonstrated to be robust against environmental noises. This high stability is achieved by letting the laser pulses to travel through the quantum channel and the QKD system twice to auto-compensate phase and polarization fluctuations. However, there are still some security concerns about this configuration. For security proofs, see [57].

(2) QKD experiments through free space

Limited by the intrinsic loss of optical fiber, it is unlikely that a fiber based QKD system can go beyond a few hundreds of kilometers. Quantum repeaters could come to the rescue. However, they are still far from practical. One major motivation behind free-space QKD is that a global quantum communication network could be built upon ground-to-satellite and satellite-to-satellite QKD links.

There are several advantages of transmitting photons in free space. First of all, the atmosphere has a high transmission window around 770 nm which coincides with the spectral range of silicon APDs. Currently, the performance of silicon APDs is superior to that of InGaAs APDs which operate at telecom wavelength (See section 3.2). Secondly, the atmosphere shows very weak dispersion and is essentially non-birefringence at the above wavelength range, which suggests that the relatively simple polarization-coding scheme can be applied. On the other hand, challenges in free space QKD are also enormous. As a laser beam propagates in free space, the beam size will increase with distance due to diffraction. To effectively collect the photon at the receiver's side, cumbersome telescopic system is required. Moreover, the atmosphere is not a static medium. Atmospheric turbulence will cause random beam wandering and sophisticated beam tracing system is required to achieve stable key transmission [41, 58].

To date, both polarization-coding decoy state BB84 QKD and entanglement based QKD have been implemented over a 144km free space link [41, 58].

3.5 Network QKD

The great potential of QKD cannot be fully appreciated until we can go beyond the point-to-point QKD and implement a QKD network. In this section, we will discuss challenges on implementing a large scale QKD network.

(1) QKD through existing fiber network

It would be much more attractive if we can implement global quantum key distribution employing the existing fiber network. However, there are several major challenges. First of all, the fiber optical amplifier, one of the most important inventions in classical optical communication, turns out to be a nightmare to QKD. A quantum state cannot be cloned; any attempts to amplify the quantum signal will spoil the encoded quantum information. Similarly, devices for signal regeneration in long haul communication links, which typically involve photo-detection process, will destroy the encoded quantum information. To protect the quantum signals, these devices have to be bypassed for QKD. This may require complicated network management.

Another challenge comes up when the quantum signals for key distribution travel together with strong classical communication signals. In classical optical communication, we can employ either time division multiplexing (TDM) or wavelength division multiplexing (WDM) to separate signals from different channels. However, the quantum signal for QKD is many orders of magnitude weaker than a classical signal. Noise photons generated by a strong classical signal through Rayleigh scattering or other nonlinear processes could be indistinguishable from the quantum signal and thus contribute to QBER.

One potential solution of the above problems is to conduct quantum communications at the 1310 nm O-band, which is generally not used for modern telecommunications. As demonstrated in [59], the classical communication signals at the 1550nm C-band have negligible impact on the quantum signal at the O-band. In the mean time, the degradation of the quantum signal after it passing through a C-band amplifier could be reduced.

(2) Large scale QKD network

QKD among multi-users in a passive local network has been proposed even in the early days of QKD [60]. Various network topologies, such as star and loop configurations, have been studied. The basic idea is that as long as a pair of users in the network establish a directly quantum link, they can perform QKD just as in the case of point-to-point QKD. However, this network architecture is not scalable. The maximum key distribution distance is still limited by the loss of the quantum channel, same as in the case of point-to-point QKD.

To extend the distance of QKD, two configurations have been proposed: quantum repeaters and trusted relays.

In classical optical communication systems, strong laser pulses also experience significant loss after passing through a long optical fiber. To extend the reach of optical communication links, "classical" repeaters have been introduced to improve the signal to noise ratio. However, as we discussed before, classical repeaters cannot be used in QKD since they typically involve optical amplification, optical to electrical conversion, electrical signal processing and optical signal regeneration.

Inspired by the concept of classical repeater, quantum repeater has been proposed to implement QKD through an arbitrary long distance [61]. Recall in the entanglement based QKD, Alice and Bob can generate secure keys as long as they can establish entanglement through a quantum channel. However, if Alice and Bob are far away from each other, they cannot establish entanglement directly due to the high channel loss. Quantum repeater solves this problem by using entanglement swapping and entanglement purification [62]. The basic idea can be understood as follows. Imagine that Charles sits in between Alice and Bob. The distance between Alice and Charles is short enough to establish entanglement between them. The same is for Bob and Charles. Once Charles shares an EPR pair $E_1$ with Alice and another EPR pair $E_2$ with Bob, he can perform a Bell-type measurement on the two half pairs in his hand and broadcast his measurement result. Based on Charles's measurement result, Alice and Bob can transform the two photons left into an EPR pair by performing local operations. Overall, by sacrificing one EPR pair, entanglement can be established over a longer distance between Alice and Bob. By employing the same scheme iteratively, entanglement, which can be used to generate secure key, can be established over an arbitrarily long distance. Note in this scheme, it is not necessary to trust Charles since he has no information about the final key.

Quantum repeaters have drawn a lot of research attention. However, a practical quantum repeater, which involves elaborated quantum operations and quantum memories, is beyond current technologies. Another approach to a scalable QKD network is based on trusted relays. This scheme can be implemented with today's technologies and has been adopted in the BBN QKD network [63] and the SECOQC QKD network [64]. Assume that two users share point-to-point QKD links with a trusted relay in the middle of them. Each user can establish a secure key with the relay using his QKD link. To establish a secure key between the two users, the trusted relay can use one time pad to encode one key with the other and broadcast the encryption results. Only the legitimate user can recover his partner's key. This scheme is scalable to arbitrary number of users and relays and thus a global QKD network can be established.

## 4. Security of practical QKD systems

Nowadays, it is not uncommon to find over-optimistic claims about QKD, such as "absolutely secure key generated by a real life QKD system", in various magazines, newspapers and even scientific journals. These claims, though some of them are not totally wrong, at least are quite misleading. First of all, the commonly used term "unconditionally secure" has a different meaning from the term "absolutely secure". As given in [11], the term "unconditional security" means that "security can be proved without imposing any restriction on the computational resources or the manipulation techniques that are available to the eavesdropper acting on the signal". Yet, conditional security may require assumptions on Alice's and Bob's devices. Secondly, what have been proved to be unconditionally secure are mathematical models of QKD systems built upon certain assumptions, rather than practical QKD systems themselves. While people can keep improving the mathematical models to get better descriptions of the actual QKD systems, a fully quantum mechanical description of a macroscopic device is not practical. In this sense, "absolutely secure" can only exist in Hilbert space.

To apply an unconditional security proof of a theoretical model to a practical QKD system, it is important to build the theoretical model upon experimentally testable assumptions and develop the corresponding measures to verify these assumptions. This is a highly nontrivial task because we have to make sure that the devices for verifying the QKD assumptions are well understood and they will not introduce additional security loopholes. Furthermore, no theoretical models can take into accounts of all the imperfections of a real-life QKD system. Is it possible that a small imperfection ignored in the security proof spoils the security of the whole system? These issues are still under investigation and we will present some recent results.

### 4.1 Verify assumptions used in a security proof

All security proofs are based on certain assumptions. It is thus very important to make sure that all these assumptions are valid in a practical QKD system. This is a two-way process: on one hand, experimentalists may need to build up sophisticated calibration devices to verify the assumptions used in the security proof; on the other hand, any assumptions which cannot be verified in practice should be removed from the QKD model.

As an example, let us review the "coherent state" assumption, which has been adopted in a well known security proof of the BB84 QKD protocol implemented with an attenuated laser source [21]. Recall that in a BB84 QKD system with a weak coherent source, due to the PNS attack, the secure key can only be generated from single-photon pulses. Thus it is important for Alice to know the photon number distribution of her source. That is where the "coherent state" assumption plays a role. Because a phase randomized coherence state has a Poissonian photon number distribution.

Although the "coherent state" assumption works well with a laser operated above its threshold, its validity is questionable in many practical QKD setups. For example, a common way to generate laser pulses is by modulating the driving current of a laser diode. The output of a laser diode during the transition period can significantly deviate from an ideal coherent state.

One way to remove this doubtful assumption has been proposed in [57]: instead of making any assumption about the photon number distribution of an "untrusted" source, Alice can perform random sampling of its output with a high speed optical switch (or a beam splitter) and a calibrated photo-detector. From the sampling results, Alice can gain enough information about the source and reestablish the unconditional security.

From the above example, it is easy to see that once an assumption has been clearly described, we can either find a way to verify it or replace it with another testable assumption. The price to pay is usually to introduce additional "calibration" or "monitoring" devices. However, the problem is still not completely solved. Because Alice and Bob have to make assumptions on how those "calibration" devices work. How do they verify these new assumptions? Should they introduce "second level calibration devices" to calibrate "first level calibration devices"? Then they may fall into an endless loop.

Obviously, they have to stop at some point by trusting some devices in their system. This highlights the fundamental difference between a QKD protocol and a real-life QKD system: while the security of the former can be proved rigorously based on quantum mechanics, the security of the latter also relies on how well we understand the QKD system.

## 4.2 Quantum hacking and countermeasures

Quantum cryptography is a competitive game between the legitimate users and the eavesdropper. If Eve always follows the rules made by Alice and Bob, she has no hope to win. On the other hand, if Eve could think differently, the game could become more interesting. For example, it has been realized that side channel attacks could be fatal in both QKD and classical cryptography. In a side channel attack, the eavesdropper tries to gain information by looking into weaknesses of the physical implementation of a cryptosystem, rather than the cryptographic algorithm itself.

In this section, we present a few examples to show how Eve can identify security loopholes in practical QKD systems by exploring imperfections ignored by Alice and Bob.

(1) Faked state attack [65]

The faked state attack is a special intercept and resent attack exploring the efficiency mismatch of the two SPDs in Bob's system. In a typical BB84 QKD setup, two separate SPDs are employed to detect bit "0" and bit "1" respectively. As we have discussed in Section 3, to reduce dark count rate, an InGaAs/InP APD-SPD is typically operated at gated Geiger mode: it is only activated for a narrow time window when Alice's laser pulse is expected. Therefore, the detection efficiency of each SPD is time dependent. In practice, due to various imperfections, the activation windows of the two SPDs may not fully overlap. Figure 6 shows an extreme case: at the expected signal arrival time $T_0$, the detection efficiencies of the two SPDs are equal. However, if the signal has been chosen to arrive at $t_0$ or $t_1$, the efficiencies of the two SPDs are substantially different.

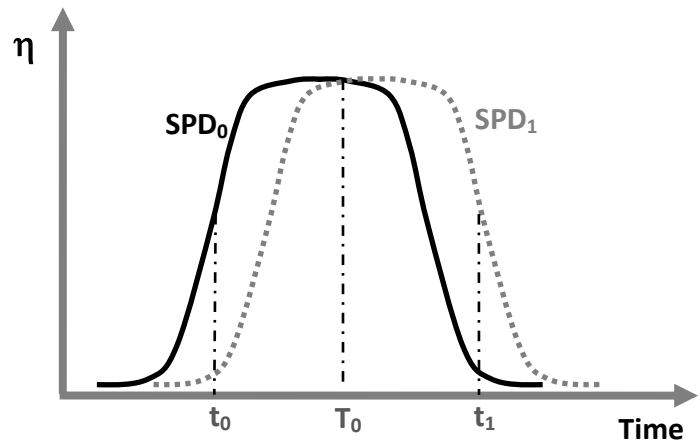

Figure 6 The efficiency mismatch of two single photon detectors in time domain. η: detection efficiency. Note at the expected signal arrival time $T_0$, the detection efficiencies of $SPD_0$ and $SPD_1$ are equal. However, if the signal has been chosen to arrive at $t_0$ or $t_1$, the efficiencies of the two SPDs are substantially different.

The faked state attack goes as follows: (1) Eve intercepts quantum states from Alice and measures each of them in a randomly chosen basis; (2) According to her measurement result, Eve prepares a new quantum state (faked state) in a different basis with a different bit value. For example, if she measures in rectilinear basis and gets bit "0", then she prepares bit "1" in diagonal basis; (3) According to her measurement result, Eve sends out her faked state at different times so that it arrives at Bob's SPD at either time $t_0$ (corresponding to her measurement result "0") or time $t_1$ (corresponding to her measurement result "1").

Eve's strategy is that whenever she uses a wrong basis, Bob's detection probability will be greatly suppressed. So she can reduce the QBER due to her attack at the cost of introducing additional loss. Note in principle, she could cover this additional loss by using a lossless channel or by sending a strong faked state to Bob.

(2) Time-shift attack [66]

The time-shift attack also exploits the efficiency mismatch of the two SPDs. The basic idea is quite simple: Eve randomly shifts the time delay of each laser pulse (but does not perform any measurement) to make sure that it arrives at Bob's SPDs at either time $t_0$ or $t_1$. If Bob registers a photon (recall that Bob will announce whether he has a detection or not for each incoming pulse), Eve will know that it is more likely from the SPD with a high efficiency and consequently she can gain partial information of Bob's bit value. In the case of a complete efficiency mismatch (which means only one of the two SPDs has a nonzero efficiency at certain time), Eve will gain as much information as Bob does thus no secure key can be generated. Time-shift attack is simple to implement because it does not require any measurement. Since Eve doesn't attempt to measure the quantum state, in principle, she will not introduce errors and cannot be detected by Alice and Bob.

It is remarkable that Eve can gain information without disturbing the quantum state. This doesn't conflict with the fundamental principles described in Section 1, which states that any attempts to gain information of an unknown quantum state will unavoidably disturb the original state. In the time-shift attack, what Eve tries to learn is the information about which SPD clicks, which is classical information. The time-shift attack can be interpreted as an "active" side channel attack: Eve encodes her own random bits on a side channel (in this case, the time shift) of the photons sent by Alice. At the beginning, there is no correlation between the random bits encoded by Alice and those by Eve. However, due to the efficiency mismatch, Bob's system has higher detection efficiency to detect photons carrying the same bits from Alice and Eve. It is this "post-selection" process introduced by Bob's detection system that correlates Eve's bits with Alice's bits, thus compromising security.

The time-shift attack can be easily implemented with today's technology and has been demonstrated on top of a modified commercial QKD system [67]. This attack can also be extended to spatial or spectral domain if the corresponding mismatch between the two detection channels exists.

Once a security loophole has been discovered, it is often not too difficult to develop countermeasures. There are several simple ways of resisting the faked state attack and time-shift attack. For example, in the phase-coding BB84 QKD system, Bob could use a four-setting set-up for his phase modulator. i.e., the phase modulator is used for not only basis selection, but also for randomly choosing which of the two SPD's is used to detect the bit "0" for each incoming pulse. Thus Eve cannot learn the bit value from the knowledge of "which detector fires". Another approach is to take into account the efficiency mismatch in the security proof, as shown in [68].

However, the difficulty remains with discovering the existence of an attack in the first place. It is, thus, very important to conduct extensive research in any implementation details of a QKD system to see if they are innocent or fatal for security. Once again, we have shown how different of the security of a practical QKD system from that of an idealized QKD protocol.

4.3 Self-testing QKD

So far we have assumed that we can trust the whole chain of QKD development, from components suppliers, system manufacturer to deliverer. This might be a reasonable assumption but in the mean time it also gives rise to a concern: is it possible that a certificated QKD manufacture leaves a trapdoor in the QKD system on purpose so that they (or some government agencies) can take this advantage to access private information? Historically, it was rumored that during the development of Data Encryption Standard (DES), National Security Agency (NSA) asked IBM to adopt a short 56 bits key for easy breaking [1].

Now, armed with quantum mechanics, can we do better? Is it possible for Alice and Bob to generate secure keys from a "black-box" QKD system which could be made by Eve?

This is the motivation behind self-testing QKD [69-70], which is also inspired by the deep connection between the security of QKD and entanglement in quantum physics.

As we have discussed in Section 2, as long as Alice and Bob can verify the existence of entanglement, it is possible to generate secure key. In the mean time, entanglement can be verified through the violation of certain Bell-type inequalities [25]. The key point is: Alice and Bob can perform Bell inequalities test without knowing how the device actually works. Imagining the QKD system as a pair of black boxes and each of the QKD users has one of them. To perform a Bell test, each user inputs a random number for basis selection to his (her) black box, which in turn outputs an experimental result. By repeating this process many times, the two users can gain enough data to test a Bell inequality. If Alice and Bob's detection results have been predetermined before they make the basis selection (for example, Eve could pre-store random bits in the two black boxes), then any observed correlation can be explained by a local hidden variable theory, which cannot violate a Bell inequality. On the other hand, if the detection results do violate a Bell inequality, these results can only be generated during the measurement processes and Eve cannot have full knowledge about them. Thus secure key generation is possible.

Another name for self-testing QKD is device-independent security proof [71], which intuitively works. However, to go from the intuition to a rigorous security proof is a very hard question. It is still an open question whether it is possible to be unconditionally secure using device-independent security proof. Furthermore, even if the unconditional security of device-independent security proofs is proven to be possible, some assumptions are still required [11]. For example, we have to trust the random number generators for choosing measurement bases in a Bell test; no information is allowed to leak out from the QKD system, etc. Anyway, assumptions used in the device-independent security proof could be much less restrictive than the ones used in the BB84 QKD protocol.

Self-testing QKD is conceptually interesting. Unfortunately, it does not yet apply to existing practical devices as a consequence of the well-known detection loophole in the experimental verification of Bell's inequalities [72]. The detection loophole states that if the detection efficiency in a Bell test is below a certain value, the observed correlation could be interpreted by a local hidden variable theory. This implies that all the detection results could be prepared by Eve ahead and no secure key can be generated.

We would like to conclude this section by remarking that not only QKD but also any classical cryptographic systems suffer from various human mistakes and imperfections of actual implementations in practice. The investigation of security loopholes and the corresponding counter-measures in practical QKD systems plays a complementary role to security proofs.

## 5. Future outlooks

25 years has passed since the publication of the celebrated "BB84" paper [8]. Thanks to the lasting efforts from physicist and engineers, QKD has evolved from a toy model operated in a laboratory into commercial products. Today, the highest secure key rate achieved is on the order of Mbit/s, which is high enough for using one time pad in certain applications [73]. At the same time, QKD over a distance of a few hundreds of kilometer has been demonstrated through both optical fiber links and free space links. Furthermore, QKD over arbitrarily long distance could be implemented by employing trusted relays.

It is very difficult to predict the future of a rather unconventional technology like QKD. In the long term, academic researchers may be more interested in fundamental issues, such as a better understanding of the security of various QKD protocols, the implementation of long distance QKD with quantum repeaters, etc. On the other hand, in the near future, one of the most urgent tasks is to integrate QKD into classical cryptographic systems and to demonstrate its usefulness in real life. Efforts by European Telecommunications Standards Institute (ETSI) to provide standardization for QKD systems are underway as we speak. The Swiss government even adopted QKD in their 2007 election [74].

Currently, there are still heated arguments about whether QKD is useful or not in practice. The central argument is not about its security or efficiency; it is more about whether QKD is solving a real problem. Quoted from Bruce Schneier, a well known security expert, "Security is a chain; it is as strong as the weakest link. Mathematical cryptography, as bad as it sometimes is, is the strongest link in most security chains". It might be true that under attacks of a classical Eve (who has no access to a quantum computer), the security of public key distribution protocol is strong. However, cryptography is a technology evolving at a fast pace. Historically, whenever a new attack appeared, the security of the old cryptographic system had to be reevaluated and quite often, people had to develop new coding scheme to reassure the security. Shor's factoring algorithm has shown that the security of many public cryptographic systems is compromised once a quantum computer is available. This is just one example that demonstrates the impact of quantum mechanics on classical cryptography. Considering the fundamental difference between quantum physics and classical physics, the security of each link in the whole chain of cryptographic system has to be carefully examined to see if it still survives in the quantum era.

QKD may not be the only solution to overcome the weaknesses of the public cryptography. People could switch back to the old cumbersome key distribution scheme by sending trusted couriers. In fact, a box of hard drives can store more random bits than a (today's) commercial QKD system could generate for years. However, the progress of a new technology opens opportunities beyond even the wildest imagination of scientists. With the joint effort of experts from different communities, perhaps it is just a matter of time that highly efficient, low-cost QKD systems are eventually realized in everyday usage.

QKD is only one of many potential applications of quantum mechanics in cryptography. Other quantum cryptographic protocols, such as quantum bit commitment and quantum coin tossing [75], etc, have also drawn a lot of attention. While it has been shown that unconditionally secure quantum bit commitment is impossible [76-77], how much advantages other quantum protocols possess over their classical counterparts are important questions under investigation.

Recently, there has been much interest in the idea of quantum cryptography based on a noisy quantum storage model. In such a model, the presumed difficulty for an adversary to store quantum signals reliably for a long time with current technology is used as a foundation of security. It has been shown that bit commitment and many other important cryptographic protocols become secure in a noisy quantum storage model. Therefore, the power of quantum cryptography is greatly enhanced. See, for example, [78] and references cited therein for details."

In a larger scope, the concept of "quantum internet" has been proposed [79]. Quantum computers, various quantum sources (such as EPR source) and quantum memories could be connected by the quantum internet, which extends its reach with the help of quantum repeaters. Although its full power is still to be seen, global QKD can be easily implemented through such a quantum network.

Quantum physics, one of the weirdest and most successful scientific theories in human history, has changed both of our view of nature and our everyday life fundamentally. The foundation stones of our information era, including transistor, integrated circuits and lasers, etc, are direct applications of quantum mechanics on a macroscopic scale. With the rise of quantum computation and quantum information, we are expecting the second quantum revolution [80].

## Acknowledgements

Support of the funding agencies CFI, CIPI, the CRC program, CIFAR, MITACS, NSERC, OIT, and QuantumWorks is gratefully acknowledged. The authors' research on QKD was supported in part by the National Science Foundation under Grant No. PHY05-51164.